# The "1–1" algorithm for Travelling Salesman Problem

Heping Jiang

jjhpjhp@gmail.com

**Abstract**
The Travelling Salesman Problem (TSP), finding a minimal weighted Hamilton cycle in a graph, is a typical problem in operation research and combinatorial optimization. In this paper, based on some novel properties on Hamilton graphs, we present a precise algorithm for finding a minimal weighted Hamilton cycle in a non-metric and symmetric graph with time complexity of $O(|E(G)|^3)$, where $|E(G)|$ is the size of graph G.

## 1. Introduction

In the Travelling Salesman Problem (TSP in abbreviation), we are given a graph such that a set of vertices along with the weighted edges on which are defined the real number values with the triangle inequality assumption (not fulfilling this assumption is called non-metric). The goal is to find a minimal weighted (sometimes denoted by distance or cost) Hamilton cycle. If the edge's weight in a graph is a constant, we call the problem a symmetric TSP.

TSP is a remarkable combinatorial optimization problem and an important subject in theoretical computer science. Solving TSP will help to impel to deal with problems in many other fields, e.g., assembly problems in bioinformatics [1-4], drilling holes in circuit broads [5], and static and dynamic vehicle routing problems [6-7] vehicle routing problems. Particularly, in solving the real-life route optimization problems, there still have the question to add driver know-how to delivery-routing models, in which many other dynamic factors that existing optimization models don't capture [8].

But, in solving these problems, the running time of precise algorithms for searching an optimal solution increases exponentially when the size of the instances grows, e.g., the dynamic programming approach to TSP [22], so the computation is intractable. With facing up the intractable computational problems, the complexity theory developed [9-10] for classifying and comparing the difficulties of solving problems about finite combinatorial objects. Henceforth, the works on TSP diverted to approximation approaches, of which most results obeyed the metric condition. In a huge amount of literature, Christofides' algorithm [11] is a landmark approximation algorithm for symmetric metric TSP. Recently, a new result reported to improve the approximation algorithm originated from Christofides' algorithm slightly [12]. But, there have no the algorithmic improvements based on exploring the existence condition of Hamilton graphs, since the problems are proved NP-hard [13]. Accordingly, in the definition of TSP, it is a crucial assumption that the given graphs are complete graphs; otherwise, it is no longer to guarantee the existence of a Hamilton cycle. In addition, most approximation algorithms are still to assume weights (distances or costs) satisfy the triangle inequality.

Unlike the conventional approximation approaches, in this paper, based on the novel properties of Hamilton graphs proposed in [14-17], we present a precise (exact) polynomial algorithm for symmetric and non-metric TSP of general graphs[1].

In [14], the Grinberg condition, a necessary condition for planar Hamilton graphs, was interpreted by using the cycle basis to replace the faces of the given graph, and then was used to derive an equation associated with inside faces called the equation of the graph. In [14], Lemma 1.1 showed the equation of a Hamilton graph has solutions, and Lemma 1.2 showed that there are two types of cycle sets in a solvable graph, (a) a 2-common ($v$, $e$) cycle set, (b) a 2-common ($v$, $0$) cycle set. Lemma 1.2 implies that we can determine the Hamiltoncity of a graph by sifting out the 2-common ($v$, $0$) cycle set, a solvable non-Hamilton graph, without planar limitation. The main notions and definitions in [14-17] appeared in this section, such as a solvable graph, etc., will be provided in Section 2.

In [15], Lemma 3.1 characterized the relation of cycle $C_k$ and the number of vertices of degree 2 in the neighbor of a vertex (denoted by $|P|$) in a solvable graph, and showed in Lemma 3.3 that a solvable norm graph is Hamilton graph, if and only if, $|C_k| = 0$.

In [16], Theorem 3.3 proved that a norm graph is non-Hamiltonian, if and only if, $\mathcal{g} \cong K_{2,3}$, where $\mathcal{g} \cong K_{2,3}$ denotes $\mathcal{g} \approx^2 K_{2,3}$ such that only one subgraph $K_{2,3}$ induced from $\mathcal{g}$.

By Lemma 3.1 and Lemma 3.3 in [15], and Theorem 3.3 in [16], for a solvable norm graph, the following statements are equivalent:

> G is non-Hamiltonian;
> $|P| \geq 3$;
> $|C_k| \neq 0$;
> $\mathcal{g} \cong K_{2,3}$.

These relations and the 2-common ($v$, $e$) cycle set are important foundations for methods to propose our new algorithm for TSP in this paper.

In the algorithm of [17], by adding the solution cycles one by one, it will construct a Hamilton cycle eventually; otherwise, it outputs the given graph is non-Hamiltonian. In a sense of R=1 edges of a graph, the algorithm in [17] continuously expands the R=1 edges in a graph for obtaining a Hamilton cycle. Using the complementary process, we continuously contract the weighted edges by deleting co-solution cycles to find a minimal weighted Hamilton cycle from a Hamilton graph, which is named the "1–1" algorithm for TSP. We show that comparing every pair of removable cycles (co-solution cycles) and deleting one only that contributes the minimal increment of weighted R=1 edge will yield a global-opt result, and repeat this process will obtain a minimal weighted Hamilton cycle. We prove that the "1–1" algorithm is a precise non-metric symmetric algorithm with time complexity of $O(|E(G)|^3)$, where $|E(G)|$ is the size of graph G.

---

[1] We use "general graphs" here means there have no restriction on complete graphs and their original graphs are simple connected graphs, whose Hamiltoncity decided by the algorithm in [17].
[2] The notation "≈" means "to be homeomorphic to" [16].

## 2. Preliminaries

We first give the main terminologies and definitions in [14-17].

Graphs considered in these papers are finite, undirected, and simple connected graphs. A graph G = (V, E) is a finite nonempty set V of elements called vertices, together with a set E of two elements subsets of V called edges. Let B(G) be a basis of the cycle space of a graph G. Two vertices u, v of G are neighbors if uv is an edge of G.

A cycle is called a removable cycle if removing it from B(G) produces a subgraph G' such that V'=V, E'=E−1 and the neighbors of any vertex in G satisfy |P|<3. R denotes the number of cycles passed through an edge in a graph G. A boundary vertex is a vertex that only has two edges of R=1 in its incident edges. A boundary edge refers to an edge of R=1. A cut vertex means a vertex that all its edges are R=1. A vertex is called inside if it is neither a boundary vertex nor a cut point.

We call $\sum(i \cdot |C_i| - 2|C_i|) = |V(G)| - 2$ the equation of a graph G, where C is the cycle with degree i in B(G). See the equation 1.2 in [14]. A solvable graph means its equation has solutions, in one of which there has a set of cycles in a basis that satisfies the equation. Every cycle in a solution is called a solution cycle; and the complement of a solution cycle is called a co-solution cycle. We denote by $S_x$ a solution of G that is a set of cycles $\{C_1, C_2, …, C_m\}$, where $1 \leq m \leq (|E(G)|-|V(G)|+1$; and $S'_x$ a co-solution that is $\{C'_1, C'_2, …, C'_n\}$, where $0 \leq n \leq (|E(G)|-|V(G)|+1) - 1$.

A 2-common (v, e) combination is a combination of cycles A and B such that $|V(A)| \cap |V(B)|=2$ and $|E(A)| \cap |E(B)|=2$; and a 2-common (v, 0) combination is a combination such that $|V(A)| \cap |V(B)|=2$ and $|E(A)| \cap |E(B)|=0$.

Let k be a boundary vertex of degree 4. $C_k$ is a cycle on which all boundary vertices are k but no edges are boundary, and $|C_k|$ is the number of $C_k$.

In a given graph, the operation of smoothing out a vertex of degree 2 consists of two processes. The first is to delete this vertex and its incident edges, and the second is to replace them by a new edge. Let |P| be the number of the vertices of degree 2 in the neighbor of a vertex. We use a reduced graph for a graph produced from deleting the edges that there have |P|=2 in the neighbor of its endpoint and smoothing out the adjacent vertices of degree 2 except for leaving one such vertex only.

Let **I** be a set of cycles jointed by a set of inside vertices adjacent each other. |**I**| is the number of sets **I** in a graph. A norm graph is a reduced graph of |**I**| = 1. $\mathscr{g}$ is a subgraph derived by deleting all removable cycles from a basis of a norm graph.

We next give the notions and definitions used in this paper.

Unless otherwise stated, graphs considered in this paper are Hamilton graphs obtained from executing the algorithm with $\mathcal{O}(|E(G)|^3)$ worst case time complexity [17]. Let G be a graph. We define the TSP to be a problem that how to find a minimal weighted Hamilton cycle in a graph G. The following definitions and notions are all involved in a graph G.

Let w denote an edge's weight, sometimes called distance or cost. There has no metric limitation for the weighted edges in a graph, which means it does not obey the triangle inequality. The edge's weight between two endpoints is constant, so we call the TSP a symmetric TSP. We use $H_{min}$ for the minimal weighted Hamilton cycle, $W_{min}$ the weight of $H_{min}$, $S_x$ a solution set, and $S'_x$ a co-solution set.

We say a cycle on the R=1 edge means it has only one R=1 edge.

Δw is an increment of the weight of R=1 edge when deleting a removable cycle. Let **t** be a

2-common (v, e) combination set in G. For x, y ∈ $g$-C we define x+**t**+y[3] is a 2-common (v, e) combination set in B(G) such that G has two components if deleting x+**t**+y. See the Figure 01.

Let $m_x$, $n_x$, a be the R=1 edges produced by deleting cycle x, and $m_y$, $n_y$, b the R=1 edges produced by deleting cycle y. Generally, we use w(mna) and w(mnb) to refer to the weights of R=1 edges generated from deleting x and y, respectively. An overlapping comparison is a comparison between w(mna) and w(mnb). A cycle is called a global-opt cycle if it is a co-solution cycle of producing the minimal weighted Hamilton cycle.

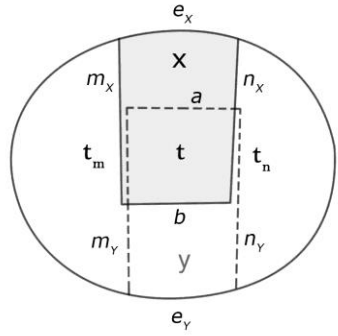

Figure 01

Let G–C be a subset of G produced by deleting cycle C from B(G). We say a cycle C is a candidate removable cycle if it satisfies |V(G–C)| = |V(G)| and |E(G–C)| = |E(G)| –1 in B(G). k is a vertex of degree 4 generated by deleting C. We denote by $C^\nabla$ a diagonal cycle to C such that $C^\nabla \cap C$ is a cut vertex (or $|E(C^\nabla)|+|E(C)|=|E(C^\nabla) \cup E(C)|$). $C^\nabla$-set is a cycle set that consists of $C^\nabla$ and the cycles associated with $C^\nabla$.

For basic terminologies of graph theory not mentioned in this paper, please refer to [18-21], others [14-17].

## 3. Methods

### 3.1 Identification of removable cycles

By the definition of removable and Lemma 3.1 in [15], for a cycle C on the R=1 edge of G, if there has no vertex K when deleting it, we can easily determine C is a removable cycle. If there has a vertex k, for identifying whether C is removable, we need first to decide whether or not there has the case $|P| \geq 3$ or $|C_k| \neq 0$ occurring in the neighbor of vertex K when deleting C.

It is easy to test the existence of $|P| \geq 3$ (actually, it was excluded in a norm graph). To identify the existence of $C_k$, by Lemma 3.3 in [15], we require asking whether or not a cycle $C^\nabla$ (which is associated with the cycle C) is a $C_k$ when deleting C. Lemma 1 states that the answer is subject to the Hamiltoncity of the $C^\nabla$-sets.

**Lemma 1** C is a removable cycle, if and only if, all $C^\nabla$-sets to C are Hamiltonian.

**Proof.** By Lemma 3.3 in [15], if all $C^\nabla$-set is Hamiltonian, then there have no $C_k$ emerge when deleting C. Note that Lemma 3.1 in [15], $|P| \geq 3 \Leftrightarrow |C_k| \neq 0$. Thus, by the definition, C is a removable cycle. On the other hand, suppose the $C^\nabla$-set is non-Hamiltonian, Then, there must

---

[3] In this paper, for cycles x and y, x+y is refer to that x connects y in the way of 2-common (v, e) combination.

have one case of the followings: $|P| \geq 3$, $|C_k| \neq 0$, or $\mathscr{G} \cong K_{2,3}$ when deleting the cycle C. By Lemma 3.1 in [15] and Lemma 3.1 in [16], we derive C is not a removable cycle. ∎

### 3.2 The Reducing of a $C^\nabla$-set

Followed Lemma 1, we then have a question that how to identify a $C^\nabla$-set's Hamiltoncity.

Note the fact that a $C^\nabla$-set is so simply a norm graph that its Hamiltoncity can be determined by a brutal method readily. Thus, we can simply use the basic rules of the Hamiltoncity in [15] to reduce a $C^\nabla$-set until we derive an acyclic subgraph or a cycle graph, the former implies that the $C^\nabla$-set is non-Hamiltonian and the later Hamiltonian. We call this procedure the $C^\nabla$-set reducing. See Lemma 2.

**Lemma 2**  A $C^\nabla$-set is Hamiltonian, if and only if, the $C^\nabla$-set reducing results in a cycle graph.

**Proof.**  Without loss of generality, we use a representative for a $C^\nabla$-set. By the definition, a $C^\nabla$-set at least has a vertex k[4], and is a $|\mathbf{I}|= 1$ [16] cycle graph. Then we can construct a representative graph that all the $C^\nabla$-sets are isomorphic to it. See the Figure 02. In the following content, we will use this representative graph to discuss the $C^\nabla$-set reducing.

We say a cycle on the R=1 edge is to express it has only one R=1 edge. Let **C** be a candidate removable cycle such that $|V((C^\nabla\text{-set})-\mathbf{C})|=|V(C^\nabla\text{-set})|$ and $|E((C^\nabla\text{-set})-\mathbf{C})|=|E(C^\nabla\text{-set})|-1$. Let $|d_2|$ denote the number of vertices of degree 2 in the neighbor of a vertex in a graph. Since the procedure we considered is the reducing, then, in the proof of this lemma, we discuss every case in terms of existence of vertices of degree 2 in the representative.

For $|d_2|=0$ in the $C^\nabla$-set, See the Figure 03, there have no need to reduce. Note that the representative is a $|\mathbf{I}|= 1$ cycle graph. Then, the vertices on $C^\nabla$ in the representative are inside vertices except for the vertex k. So, if we delete the cycle on the R=1 edge of the

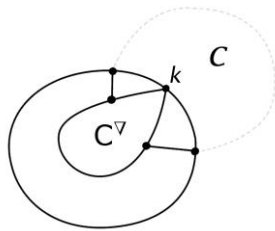
Figure 02  The representative of a $C^\nabla$-set

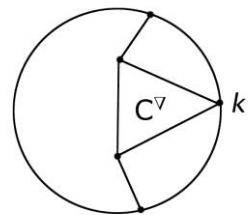
Figure 03  $|d_2|=0$

representative, denoted by C', as it satisfies $|V((C^\nabla\text{-set}) -C')| = |V(C^\nabla\text{-set})|$ and $|E((C^\nabla\text{-set}) -C')| = |E(C^\nabla\text{-set})|-1$, we will change all inside vertices boundary, which implies we obtain a 2-common (v, e) combination set of the $C^\nabla$-set. This is a cycle graph and is Hamiltonian.

If $|d_2|=1$, then we have five cases of the $C^\nabla$-set. See the Figure 04. All of them have no need to reduce. Similar to the case $|d_2|=0$, we also derive that the $C^\nabla$-set is Hamiltonian because that it is also a cycle graph.

---

[4] Using "at least" here actually means that we can use one vertex k to represent others. Since no matter how many vertices k on cycle C, it is the same procedure to reduce the $C^\nabla$-set for every vertex k.

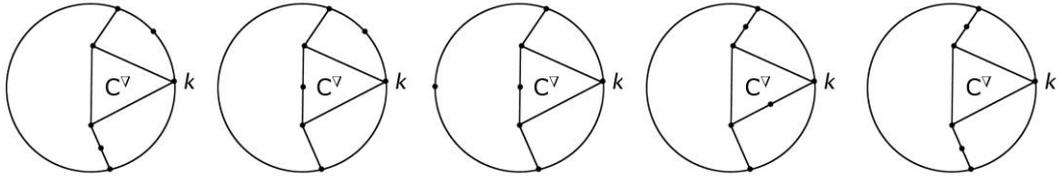

Figure 04　The five cases of $|d_2|=1$ in the representative of a $C^\nabla$-set

If $|d_2|=2$, then there have two subcases: (i) at least one cycle on the R=1 edge of $C^\nabla$-set is a cycle C′, see the Figure 05(a); (ii) no cycle on the R=1 edge of $C^\nabla$-set is a cycle C′, see the Figure 05(b).

For subcase (i), by the reducing, we derive a cycle graph that is a 2-common (v, e) combination set of the $C^\nabla$-set, which is also a cycle graph. Thus, the $C^\nabla$-set is Hamiltonian. For subcase (ii), it is clearly that reducing the $C^\nabla$-set will result in an acyclic graph. In this graph, we find a case of $|P|\geq 3$, By Lemma 3.3 in [15], the $C^\nabla$-set is non-Hamiltonian.

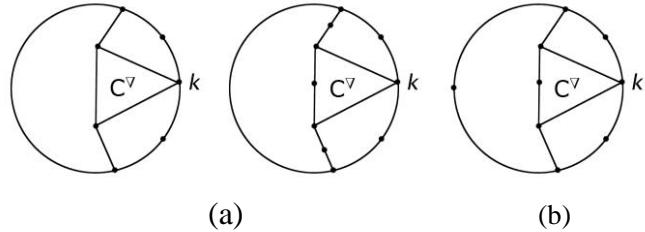

(a)　　　　　　(b)

Figure 05　$|d_2|=2$ in the representative of a $C^\nabla$-set

Finally, we consider $|d_2|=3$. Note that G is a norm graph. Since G has no case of $|P|\geq 3$, then the case of $|d_2|=3$ can only emerge when deleting the cycle **C**. It must be respect to the vertex k. The result of following process is same as that of subcase (ii) in $|d_2|=2$. We complete the proof. ∎

### 3.3　The overlapping comparison

**Lemma 3**　The cycle with the minimal Δw in the overlapping comparison is a unique global-opt cycle.

**Proof.**　Note that G is a Hamilton graph. So, G must have a co-solution cycle set S′$_x$ whom is complementary to a solution cycle set S$_x$ except that G itself is a 2-common (v, e) combination set. Then, there must have removable co-solution cycles on the R=1 edge of B(G). Thus, we have to compare them for deciding which one is the best choice to delete for obtaining a minimal weighted Hamilton cycle.

According to the definition, S$_x$ and S′$_x$ are complementary. Let sub-S′$_x$ be a subset of S′$_x$ such that the union of all the sub-S′$_x$ is complementary to S$_x$. Let C′$_x$ be the cycles in S′$_x$. Let |sub-S′$_x$| denote the number of subsets in S′$_x$ and |C′$_x$| the number of cycles in a sub-S′$_x$. In an overlapping comparison, we assume cycle y is global-opt. Cycles considered in this proof are removable, unless otherwise stated.

Clearly, if |sub-S′$_x$|=1 and |C′$_x$|≥1, then we have only one cycle on the R=1 edge of B(G) to be cycle y. Obviously, cycle y is the unique and contributes the minimal Δw. This means there

have no need for the overlapping comparison, and then cycle y is global-opt. See the Figure 06 (a).

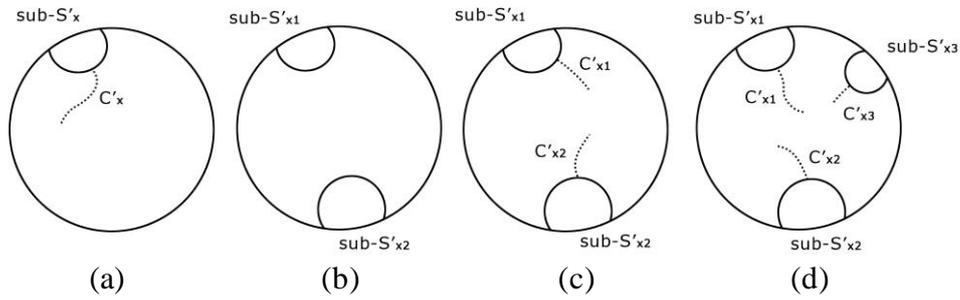

Figure 06    The cases of sub-S′$_x$ and C′$_x$. Dot lines are refer to the C′$_x$

If |sub-S′$_x$|=2 and | C′$_x$ |=1, clearly, any cycle to be y in the overlapping comparison that gives the minimal Δw is global-opt. See the Figure 06 (b).

If |sub-S′$_x$|=2 and | C′$_x$ |>1, see the Figure 06 (c). By the definition of sub-S′$_x$, the sum of all the cycles in a sub-S′$_x$ is a cycle with the same order of sub-S′$_x$. This is equivalent to the situation of |sub-S′$_x$|=2 and |C′$_x$|=1, so any cycle involving in a sub-S′$_x$ to be y in the overlapping comparison is also global-opt.

For the case of |sub-S′$_x$|≥3 and |C′$_x$|≥1, see the Figure 06 (d). This situation is an expanded version of the cases (b) and (c). Then, we are easily to derive a global-opt cycle y.

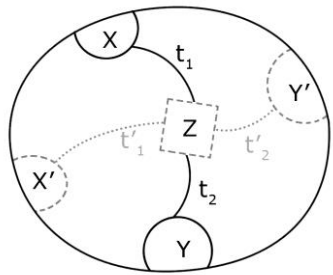

Figure 07    $t_1$, $t_2$, $t_1'$, and $t_2'$ are refer to the C′$_x$

Now we consider that if there is a cycle not lying on the R=1 edge of B(G), denoted by z, such that it would be a global-opt cycle only when it is in a different combination, e.g., x′ + $t_1'$+ z + $t_2'$ + y′. See the Figure 07. Suppose there is a global-opt cycle z in an overlapping comparison of G. As a different combination, it must has at least one cycle (x′ or y′) lying on the R=1 edge of B(G) and must be an optimal cycle. Clearly, cycle x′ or y′ must have been tested in overlapping comparison among x, x′, y, and y′, and we should derive that selecting cycle y is optimal. This contradicts to that x′ or y′ is an optimal cycle. Then, the combination of x′ + $t_1'$+ z + $t_2'$ + y′ does not an optimal selection. Hence, z is not a global-opt cycle in the combination of x′ + $t_1'$+ z + $t_2'$ + y′. This indicate that cycle y as a global-opt cycle is unique.

That completes the proof. ■

## 4. The "1–1" algorithm theorem

Combined the methods of Identification of removable cycles, $C^\nabla$-set Reducing (for identifying the removable co-solution cycles), and overlapping comparison, we obtain a new precise algorithm for TSP.

**Theorem 4** The "1–1" algorithm is an $O(|E(G)|^3)$ worst-case time algorithm for TSP.

**Proof.** According to the definition of TSP in this paper, the proof has two parts. The algorithm results in $H_{min}$ and has the worst time complexity of $O(|E(G)|^3)$.

**Part I** The "1–1" algorithm results in $H_{min}$.
For a given Hamilton graph G obtained by the algorithm in [17], we have the following inputs: $B(G)$, $S'_x = \{C_1, C_2, \ldots, C_x\}$, $g$-$C = \{C_1, C_2, \ldots, C_y\}$. We use the cycle matrix $\mathbf{M}_B = (a_{ij})_{|B(G)| \times |E(G)|}$ for $B(G)$, and $\mathbf{M}_{S'_x} = (a_{ij})_{|S'_x| \times |E(G)|}$ for $S'_x$. All the computations in the program will be based on the cycle matrix.

Executing three main steps below will output the $H_{min}$.

S1 Identify the removable co-solution cycles on the R=1 edge.
S2 Do overlapping comparison between x and y and deleting one with $\Delta w$.
S3 Repeat S1~S2 until no cycle can be selected.

Lemma 1 and Lemma 2 give us the method to identify the candidate removable cycles on the R=1 edge of $B_{min}$ and its subsets, which guarantees all the selected co-solution cycles are removable. By Lemma 3, we obtain a global-opt cycle with the minimal $\Delta w$ in the overlapping comparison.

Then, if the $\Delta w$ of all the co-solution cycles have been evaluated in the overlapping comparison, the program derives a result that no cycles can be selected to construct a more optimal weighted Hamilton cycle. Hence, the program terminates with the minimal weighted Hamilton cycle of G.

**Part II** The "1–1" algorithm has the worst time complexity of $O(|E(G)|^3)$.
In S1, for finding a global-opt removable cycle, we need to identify the cycle is removable. First, we use the number of general co-solution cycles $|g$-$C|$ to represent the number of candidate cycles. Clearly, $|g$-$C| \leq |E(G)| - |V(G)| + 1$[5]. So, $|g$-$C| \leq |E(G)|$, and there have no more $|E(G)|$ candidate cycles for the comparison. Accordingly, we have $|S'_x| = |E(G)|$ steps for computation in the program. Second, we need to identify that the candidate cycle is removable. It costs $5|E(G)|^2$ times of matrix computation, which involve finding and identifying it is a cycle in $|g$-$C|$, finding a $C^\nabla$, constructing the $C^\nabla$-set, and reducing the $C^\nabla$-set to decide whether it is Hamiltonian or not.

In S2, we first calculate the $\Delta w$ of the candidate cycle by the matrix computation, which needs $|E(G)|$ times procedures. And then we compare the $\Delta w$ to that of another candidate cycle. Therefore, the total cost is $2|E(G)|^2$.

By the definition of overlapping comparison, in the next step, we delete the cycle with smaller $\Delta w$ and repeat the next comparison. Note that we have at most $|E(G)|$ candidate cycles. Therefore, the times of implement of the whole overlapping comparison is $|E(G)| \cdot (5|E(G)|^2 + 2|E(G)|^2)$. Thus, we derive the time complexity is $O(|E(G)|^3)$, where $|E(G)|$ is the size of graph G. We complete the proof. ∎

---

[5] If $|g$-$C| = |E(G)| - |V(G)| + 1$, then G has only one Hamilton cycle.

## 5. Conclusions and discussions

In this paper, based on the works in [14-17], we develop the methods of $C^\nabla$-set reducing and overlapping comparison, and present a new non-metric precise algorithm for finding a minimal weighted Hamilton cycle with time complexity of $O(|E(G)|^3)$. Our algorithm has some distinctive features that are correspondent to theoretical and practical problems.

First, the input graphs are normalized and Hamilton graphs which identified by the algorithm in $\mathcal{O}(|E(G)|^3)$ worst case time complexity [17]. This feature implies our algorithm has no limitation in the definition of the TSP that the given graph is a complete graph, whereas most of the state-of-art approximate algorithms need the assumption to guarantee their algorithms obtain a correct result. A typical instance is on optimal walking tour through thousands cities. In solving the problem, it must first determine that the graph of these cities connecting with roads geographically is Hamiltonian; otherwise, we may intend to search a minimal weighted Hamilton cycle in a non-Hamilton graph.

Second, the crucial feature of our algorithm is that in the comparison we can omit to evaluate the weights of the common interval ($t_m+t_n$ in the Figure 01) in the cycle set because this interval is a 2-common (v, e) combination set. This property provides a very efficient searching method that significantly decreases the obstacle emerged in the precise algorithms for TSP.

The third predominant feature is that the "1–1" algorithm is non-metric. Since we only compute and compare the increments generated from deleting a pair of candidate removable cycles, then the given weights (costs or distances) for edges do not need to obey the triangle inequality. This allows the weighted value can be substituted by the data collected from various dynamic factors. For instance, we can use the real-life data [8] that encode the driver's know-how in the route optimization problems, which is an optimization problem comes from TSP.

## References

[1] John Gallant, On Finding Minimal Length Superstrings, Journal of computer and system sciences 20, 50-58 (1980)  https://doi.org/10.1016/0022-0000(80)90004-5

[2] John Gallant, The complexity of the overlap method for sequencing biopolymers, Journal of Theoretical biology 101, 1-17, 1983 https://doi.org/10.1016/0022-5193(83)90270-9

[3] J. Kececioglu, D. Sankoff, Exact and Approximation Algorithms for Sorting by Reversals, with Application to Genome Rearrangement, Algorithmica (1995) 13: 180-210. https://doi.org/10.1007/BF01188586

[4] J. Kececioglu, E. W. Myers, Combinatorical algorithms for DNA sequence assembly, Algorithmica (1995) 13: 13-51. https://doi.org/10.1007/BF01188580

[5] Grötschel, M., M. Jünger, G. Reinelt, Optimal Control of Plotting and Drilling Machines: A Case Study, Mathematical Methods of Operations Research, Vol. 35, No. 1, (January, 1991), pp.61-84. https://doi.org/10.1007/BF01415960

[6] Lenstra, J.K,., Rinnooy Kan, A.H.G. (1975). Some simple applications of the traveling salesman problem. Operational Research Quarterly, Vol. 26, pp. 717–733. https://doi.org/10.2307/3008306

[7] B.G. Madsen, A. Larsen, M. M. Solomon, *Dynamic Vehicle Routing Systems – Survey and Classification,* Proceeding of the Tristan IV Conference (2007)

[8]https://www.amazon.science/blog/amazon-mit-team-up-to-add-driver-know-how-to-delivery-r

outing-models

[9] Cook, S. A. (1971), The complexity of theorem-proving procedures, Proc. 3rd Ann. ACM symp. On Theory of Computing, Association for computing Machinery, New York, 151-158. https://doi.org/10.1145/800157.805047

[10] Richard M Karp. Reducibility among combinatorial problems. Springer, Boston, MA, 1972. https://doi.org/10.1007/978-1-4684-2001-2_9

[11] Nicos Choristofides, Worst Case Analysis of a New Heuristic for the Travelling Salesman Problem, Technical Report 388, Graduate School of Industrial Administration, Carnegie-Mellon University,Pittsburgh, Pennsylvania. 1976

[12] Anna R. Karlin, Nathan Klein, Shayan Oveis Gharan, An Improved Approximation Algorithm for TSP in the Half Integral Case,

[13] Michael R. Garey and David S. Johnson. Computers and Intractability; A Guide to the Theory of NP-Completeness. W. H. Freeman & Co., USA, 1990.

[14] A note on the Grinberg condition in the cycle spaces https://arxiv.org/abs/1807.10187v6

[15] Non-Hamiltonian cycle sets of having solutions and their properties
https://arxiv.org/abs/1906.09678v2

[16] The induced subgraph $K_{2,3}$ in a Non-Hamiltonian graph
https://arxiv.org/abs/1906.10847v2

[17] A "1+1" algorithm for the Hamilton Cycle
https://arxiv.org/abs/1710.06974v2

[18] C. Berge. Graphs, Third revised ed. Elsevier Science Publishing Company,INC, 1991.

[19] J.A.Bondy and U.S.R.Murty. Graph Theory. Springer, 2008.

[20] Narsingh Deo. Graph Theory with Applications to Engineering and Computer Science. Prentice Hall, Inc., Englewood Cliffs, N. J., 1974.

[21] Reinhard Diestel. Graph Theory. Springer-Verlag, New York, 2000.

[22] Richard Bellman. Dynamic programming treatment of the travelling salesman problem. J. ACM, 9 (1), 1962, 61-63. https://doi.org/10.1145/321105.321111